# Depolarization of Synchrotron Polarization in Magnetized Interstellar Media

Zheng-Kang Yang,[1] Jian-Fu Zhang,[1,2] Zi-Qi Liu,[1] and Qing-Zhong Liu[1]

[1] *Department of Physics, Xiangtan University, Xiangtan, Hunan 411105, People's Republic of China;*
[2] *National Astronomical Observatories, Chinese Academy of Sciences, 20A Datun Road, Chaoyang District, Beijing 100012, China*

## ABSTRACT

We numerically study three key depolarization mechanisms of synchrotron polarization, including wavelength-independent depolarization, differential Faraday rotation depolarization, and beam depolarization, across sub-Alfvénic and super-Alfvénic turbulent regimes. Using 3D MHD simulations and synthetic observations, we analyze their effects on polarization degree and power spectra. We find that (1) Sub-Alfvénic turbulence shows high but strongly suppressed polarization under Faraday rotation, while super-Alfvénic turbulence remains low and stable; (2) Differential Faraday depolarization dominates in ordered sub-Alfvénic media. For disordered super-Alfvénic regimes, beam depolarization dominates at high frequencies, while at low frequencies it saturates and differential Faraday depolarization takes over as the dominant mechanism; and (3) Depolarization significantly distorts spectra in super-Alfvénic turbulence but weakly affects sub-Alfvénic cases. Our results provide a quantitative basis for correcting depolarization biases in magnetic turbulence measurements with radio polarization data.

## 1. INTRODUCTION

Turbulence is ubiquitous in key astrophysical environments (A. Chepurnov & A. Lazarian 2010), most of which are magnetized and host relativistic electrons, making synchrotron emission a valuable probe for astrophysical magnetic fields. Given that the polarization properties of synchrotron radiation are closely associated with magnetic field characteristics, analyzing polarization features can facilitate our understanding of magnetic field properties (A. H. Waelkens et al. 2009; A. Lazarian & D. Pogosyan 2012, 2016). A correct understanding of these magnetic field properties plays a pivotal role in investigating fundamental astrophysical processes, such as star formation and cosmic ray propagation (see B. Burkhart 2021 for a comprehensive review).

Random fluctuations in magnetic fields naturally induce fluctuations in the observed synchrotron polarization radiation. These fluctuations introduce significant foreground contamination, posing substantial challenges for high-precision measurements of the cosmic microwave background (CMB; A. J. Banday & A. W. Wolfendale 1991; N. Krachmalnicoff et al. 2018) and for studies of fluctuations in the atomic hydrogen distribution in the early universe (A. Ghosh et al. 2012). Thus, a comprehensive understanding and rigorous analysis of turbulence characteristics are crucial for separating these contaminating signals from the target astrophysical signals (J. Cho & A. Lazarian 2010).

Notably, intrinsic fluctuations in turbulent magnetic fields govern synchrotron depolarization (A. Kogut & D. C. Collins 2025), while Faraday rotation (FR) introduces additional polarization fluctuations by rotating the polarization plane in an angle proportional to $\lambda^2 \int n_e B_{\|} dz$. Collectively, these effects make depolarization an inherent observational signature of magnetic turbulence (B. J. Burn 1966; D. D. Sokoloff et al. 1998). Synchrotron depolarization effects give rise to systematic discrepancies between the measured polarization signals and the intrinsic polarization signals revealing local physical conditions of the emission sources (H. Junklewitz & T. A. Enßlin 2011), and they remain challenging to disentangle in radio polarization observations. The core barrier to this disentanglement lies in the fact that real observations always capture the combined signal from multiple depolarization processes such as wavelength-independent depolarization, differ-

Email: jfzhang@xtu.edu.cn (JFZ), qzliu@pmo.ac.cn

ential FR depolarization, and beam width depolarization.

In multi-frequency observations, the application of traditional rotation measure (RM) synthesis (M. A. Brentjens & A. G. de Bruyn 2005; see also B. J. Burn 1966 for the original version) is constrained by the frequency sampling interval and total bandwidth, which introduces ambiguities when interpreting Faraday-complex media containing turbulent line-of-sight (LOS) components (K. Bekki 2023). While high-resolution interferometry can resolve fine-scale structures, it sacrifices sensitivity to large-scale polarization in extended sources and fails to address intrinsic LOS fluctuations. Additionally, narrow band observations mitigate band width depolarization but reduce signal strength and the spectral signal-to-noise ratio, thereby hindering the detection of weak polarized signals.

Previous studies have shown that various turbulence measurement techniques established on synchrotron polarization fluctuation theory (A. Lazarian & D. Pogosyan 2012, 2016), including the gradient technique (see J.-F. Zhang et al. 2019 for measuring magnetic field direction, L. Carmo et al. 2020 for magnetic field strength, and J.-F. Zhang & Z.-Q. Liu 2025 for power-law slope), the quadrupole ratio modulus technique (R.-Y. Wang et al. 2020 for measuring anisotropy of MHD turbulence), the polarization variance technique (J.-F. Zhang et al. 2016, Y.-W. Xiao et al. 2025, Y.-W. Xiao et al. 2026 for magnetic field power-law spectrum and N.-N. Guo et al. 2024 for the magnetization strength), and the power spectrum method (J.-F. Zhang et al. 2018 for the scaling slope of the magnetic field), are all constrained to a certain extent by FR depolarization. A conclusion commonly drawn from these studies is that the measurement of the gradient technique is almost unaffected by FR depolarization caused by wavelength variations; however, it remains unclear how the other depolarization methods mentioned above affect the application of polarization analysis techniques in measuring magnetic turbulence properties.

Motivated by the limitations of existing magnetic turbulence measurement techniques and the unresolved question of how most depolarization methods except for the minimal impact of FR depolarization on gradient-based techniques affect magnetic turbulence measurement methods, our work aims to decouple the influences of various depolarization processes on these measurement techniques (focusing on polarization degree and power spectrum) across different turbulence regimes. We outline the theoretical framework of depolarization in Sect. 2, detail the numerical setup of our simulations in Sect. 3, present our key results in Sect. 4, followed by a discussion in Sect. 5 and a summary in Sect. 6.

## 2. THEORETICAL FOUNDATIONS OF DEPOLARIZATION

### 2.1. *Polarization from Synchrotron Emission*

Polarized synchrotron emission is produced by relativistic electrons gyrating in magnetized media. The energy distribution of relativistic electrons is commonly assumed to follow a power-law form:

$$N(E)\,dE = N_0 E^{-\gamma}\,dE, \tag{1}$$

where $E$ denotes the electron energy, $N(E)$ is the number density of electrons per unit energy interval, $N_0$ is a normalization constant, and $\gamma$ is the spectral index.

For a synchrotron-emitting region extended along the $z$-axis (the LOS direction), the observed polarized emission results from the line-of-sight integration of local synchrotron emissivity. The linear polarization is characterized by the Stokes parameters, which can be written as (e.g., A. H. Waelkens et al. 2009)

$$Q_0 \propto \nu^{\frac{1-\gamma}{2}} \int_0^L \left(B_x^2 + B_y^2\right)^{\frac{\gamma-3}{4}} \left(B_x^2 - B_y^2\right) dz, \tag{2}$$

$$U_0 \propto \nu^{\frac{1-\gamma}{2}} \int_0^L \left(B_x^2 + B_y^2\right)^{\frac{\gamma-3}{4}} \left(2B_x B_y\right) dz, \tag{3}$$

where $L$ is the integration depth along the LOS. The synchrotron total intensity is correspondingly given by

$$I \propto \nu^{\frac{1-\gamma}{2}} \int_0^L \left(B_x^2 + B_y^2\right)^{\frac{\gamma-3}{4}} \left(B_x^2 + B_y^2\right) dz. \tag{4}$$

Defining the complex polarization vector as $P = Q_0 + iU_0$, we can write the observed degree of polarization as [3]

$$p_\alpha = \frac{|P|}{I} = \frac{\sqrt{Q_0^2 + U_0^2}}{I}. \tag{5}$$

### 2.2. *Wavelength-Independent Depolarization*

In the ideal case of a perfectly ordered magnetic field, the observed polarization degree approaches the intrinsic synchrotron polarization fraction $p_0$. However, in realistic magnetized astrophysical media, various physical and observational effects reduce the coherence of polarized emission, leading to depolarization of the observed synchrotron signal.

[3] For a power-law electron energy spectrum with $\gamma = 2.5$, the maximum intrinsic synchrotron polarization fraction in a perfectly ordered magnetic field is $\Pi_{\max} = (\gamma+1)/(\gamma+7/3) \approx 0.724$ (B. J. Burn 1966). This factor is used to scale the intrinsic polarization degree in our calculations.

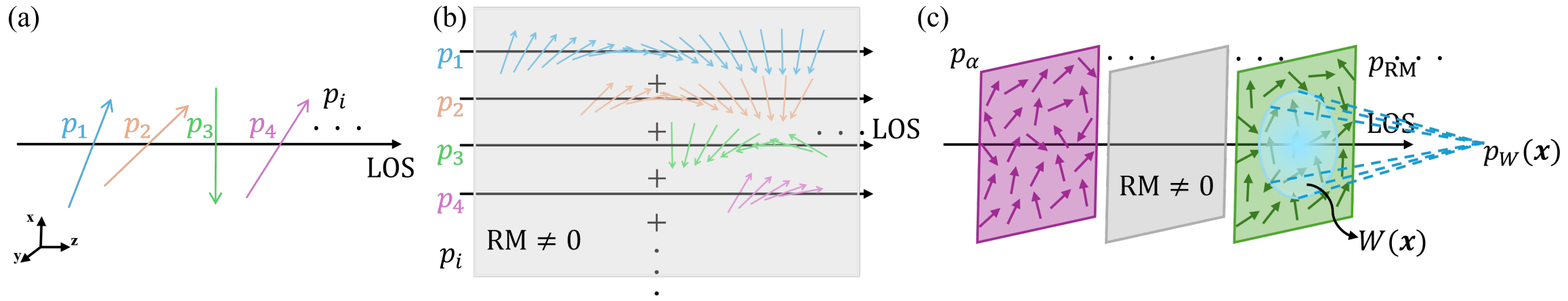

**Figure 1.** Schematic diagram of depolarization mechanisms: wavelength-independent depolarization (panel (a)), differential Faraday rotation depolarization (panel (b)), and beam depolarization (panel (c)). The LOS is oriented along the $z$-axis. The vector symbols $p_i$ ($i = 1, 2, 3, \ldots$) represent the polarization directions on the plane of the sky. The colored polarization vectors in panel (b) correspond to the same emitting regions shown in panel (a). In panel (c), $p_\alpha$ denotes the polarization degree after wavelength-independent depolarization, $p_{\rm RM}$ represents the polarization degree after Faraday rotation depolarization, and $p_{\rm W}$ is the polarization degree after beam depolarization. The function $W(\mathbf{x})$ represents the telescope beam profile.

Wavelength-independent depolarization arises from random fluctuations in the orientation of magnetic fields perpendicular to the LOS within turbulent magnetized media (see Fig. 1(a)). This effect results from the incoherent superposition of polarized synchrotron emission produced at different positions along the LOS (T. Akahori et al. 2018), leading to partial cancellation of polarization vectors, and consequently decreases the observed polarization degree below the intrinsic value $p_0$.

### 2.3. *Differential Faraday Rotation Depolarization*

Even when the magnetic field perpendicular to the LOS remains spatially coherent, depolarization can still occur in the presence of a nonzero RM along the LOS. In magnetized thermal plasmas, polarized synchrotron emission undergoes Faraday rotation during propagation through the medium, causing the polarization angle to evolve with wavelength and LOS depth (see Fig. 1(b)).

The polarization angle after Faraday rotation is given by

$$\chi(z, \lambda^2) = \chi_0 + {\rm RM}(z)\lambda^2, \tag{6}$$

where $\chi_0$ is the intrinsic polarization angle of the synchrotron radiation, $\lambda$ is the observed wavelength, and ${\rm RM}(z)$ is the Faraday depth defined as

$$\begin{aligned}{\rm RM}(z) &= \frac{e^3}{2\pi m^2 c^4}\int_0^z n_e B_\parallel ds \\ &= \int_0^z \left(\frac{n_e}{\rm cm^{-3}}\right)\left(\frac{B_\parallel}{1.23\ \mu{\rm G}}\right) d\left(\frac{s}{\rm pc}\right)\ {\rm rad\,m^{-2}},\end{aligned} \tag{7}$$

where $n_e$ is the thermal electron number density and $B_\parallel$ is the magnetic field component parallel to the LOS. When adopting a normalized form, the constant factor $\sim 0.81$ in Eq. (7) is absorbed into the normalization.

When polarized emission originating from different depths undergoes different amounts of Faraday rotation, the resulting polarization vectors acquire differential phase shifts and become partially incoherent. The associated vector cancellation reduces the net observed polarization, producing differential Faraday rotation depolarization (T. Akahori et al. 2018). In differential Faraday rotation depolarization, the Stokes parameters are modified by Faraday rotation as

$$Q(\lambda^2) = Q_0 \cos 2\chi + U_0 \sin 2\chi, \tag{8}$$

$$U(\lambda^2) = U_0 \cos 2\chi - Q_0 \sin 2\chi, \tag{9}$$

where $\chi$ is the wavelength-dependent polarization angle defined by Equation (6). Consequently, the observed polarization degree after Faraday depolarization is then given by

$$p_{\rm RM}(\lambda^2) = \frac{\sqrt{Q^2(\lambda^2) + U^2(\lambda^2)}}{I}. \tag{10}$$

### 2.4. *Beam Depolarization*

Beam depolarization arises from the finite angular resolution of radio telescopes (see Fig. 1(c)). When polarized emission contains unresolved spatial fluctuations within the telescope beam, polarization vectors originating from different regions are averaged together, leading to partial vector cancellation and a reduction in the observed polarization degree (T. Akahori et al. 2018; P. C. Tribble 1991). The strength of beam depolarization therefore depends on both the spatial structure of the polarized emission and the telescope beam size.

For a telescope with beam function $W(\mathbf{x})$, where $\mathbf{x}$ denotes celestial coordinates on the plane of the sky, the observed complex polarization corresponds to the convolution of the intrinsic polarization distribution with the telescope beam:

$$\begin{aligned}P(\mathbf{x}_0, \lambda^2) = \int & W(\mathbf{x}_0 - \mathbf{x})\, e(\mathbf{x}) \\ & \times \exp\left[2i{\rm RM}(\mathbf{x})\lambda^2 + 2i\theta(\mathbf{x})\right] d^2\mathbf{x},\end{aligned} \tag{11}$$

where $e(\mathbf{x})$ denotes the intrinsic polarized intensity and $\theta(\mathbf{x})$ represents the initial position angle.

In turbulent magnetized media, beam depolarization is commonly described statistically through the ensemble average of the squared polarization amplitude (P. C. Tribble 1991):

$$\langle|P_{\rm W}|^2\rangle = \int d^2\mathbf{x} \int d^2\mathbf{x}'\ R(\mathbf{x}-\mathbf{x}')\, e(\mathbf{x})\, e(\mathbf{x}') \\ \times \exp\left[2i\,{\rm RM}(\mathbf{x})\lambda^2 - 2i\,{\rm RM}(\mathbf{x}')\lambda^2\right] \\ \times \exp\left[2i\theta(\mathbf{x}) - 2i\theta(\mathbf{x}')\right], \tag{12}$$

where $R(\mathbf{s}) = \int d^2\mathbf{x}\, W(\mathbf{x})\, W(\mathbf{x}+\mathbf{s})$ is the spatial correlation function of the telescope beam. The corresponding observed polarization degree can be expressed as

$$p_{\rm W} = \frac{\sqrt{\langle|P_{\rm W}|^2\rangle}}{I}, \tag{13}$$

where $I$ is the total synchrotron intensity. Beam averaging over unresolved polarization fluctuations reduces $\langle|P_{\rm W}|^2\rangle$ and consequently decreases the observed polarization degree.

For radio astronomical observations, the telescope beam is commonly approximated by a Gaussian profile:

$$W(\mathbf{x}) = \frac{1}{\pi t^2} \exp\left(-|\mathbf{x}|^2/t^2\right), \tag{14}$$

where $t$ is the characteristic scale of the Gaussian beam, and the full width at half maximum (FWHM) $w$ satisfies $w = 2t\sqrt{\ln 2}$. The corresponding beam spatial correlation function is

$$R(\mathbf{s}) = \frac{1}{2\pi t^2} \exp\left(-|\mathbf{s}|^2/2t^2\right). \tag{15}$$

Larger beam sizes increase spatial averaging over polarization fluctuations and therefore enhance beam depolarization.

## 3. GENERATION OF SIMULATION DATA

### 3.1. *Description of MHD Simulations and Related Parameters*

In this work, we perform 3D compressible MHD turbulence simulations for the diffuse interstellar medium using the second-order accurate hybrid essentially non-oscillatory MHD code developed by J. Cho & A. Lazarian (2003). Within a single-fluid proton framework, we solve the full set of MHD equations adopting an isothermal equation of state to follow the evolution of turbulent magnetic fields, gas velocity, and density distributions.

A uniform background magnetic field $B_0$ is initialized along the $x$-axis, such that the total magnetic field reads $\boldsymbol{B} = \boldsymbol{B}_0 + \delta\boldsymbol{B}$, where $\delta\boldsymbol{B}$ denotes the turbulent fluctuating component. Turbulence is driven by large-scale divergence-free random forcing applied in Fourier space at wavenumber $k \simeq 2.5$. Taking the initial gas pressure as a tunable parameter under the above configuration, we carry out simulations on a periodic Cartesian domain of side length $2\pi$ with a numerical resolution of $512^3$. After the simulations attain a statistically steady state, we output snapshots of the three-dimensional velocity, magnetic field, and mass density $\rho$. Using these snapshots, we compute secondary diagnostics summarized in Table 1, including mode-energy ratios, the Alfvénic Mach number $M_{\rm A} = \langle|\boldsymbol{v}|/V_{\rm A}\rangle$, the sonic Mach number $M_{\rm s} = \langle|\boldsymbol{v}|/c_{\rm s}\rangle$, the plasma beta parameter $\beta = 2M_{\rm A}^2/M_{\rm s}^2$, and spectral indices. Here $\boldsymbol{v}$ is the turbulent velocity, $V_{\rm A}$ the Alfvén velocity, and $c_{\rm s} = \sqrt{P_0/\rho}$ the sound speed. These diagnostics provide good characterizations for the properties of individual turbulence regimes.

Furthermore, to investigate how magnetic turbulence strength impacts distinct depolarization mechanisms, we rotate the simulation data cubes corresponding to the four turbulence regimes listed in Table 1 (see Appendix A for our rotation procedure). Every cube undergoes 50 independent rotations, which produce an ensemble of 50 realizations spanning a range of mean parallel $\langle B_\parallel\rangle$ and mean perpendicular $\langle B_\perp\rangle$ magnetic fields.

### 3.2. *Generation of Synthetic Observations*

To generate synthetic radio polarization observations, the dimensionless outputs of the MHD simulations are scaled to physical quantities representative of typical conditions in the diffuse Galactic interstellar medium. Each magnetic field cube is normalized by its volume-averaged mean and uniformly scaled to a physical mean strength of $B \simeq 0.123\ \mu$G (R. Jansson & G. R. Farrar 2012; A. R. Taylor et al. 2009). We adopt a thermal electron density of $n_{\rm e} \simeq 0.01$ cm$^{-3}$ and a LOS integration depth of $L = 1$ kpc, corresponding to the physical size of the simulation domain (S. K. Ocker et al. 2020).

For synchrotron emission calculations, the relativistic electron population is assumed to follow the power-law energy distribution given in Eq. (1), with a fixed spectral index of $\gamma = 2.5$ adopted throughout this work.[4] Following the theoretical framework described in Sect. 2, we compute the polarized intensity and polarization degree for different depolarization mechanisms. For beam depolarization, we consider the simplified configuration in which the synchrotron-emitting region and the Faraday-

[4] Relativistic electrons intrinsically produce a synchrotron spectral scaling $\nu^{(1-\gamma)/2}$. Following LP16, we omit this scaling when computing polarization statistics, as it only rescales the amplitude of statistic quantities without modifying the underlying statistical properties. This is numerically supported by J.-F. Zhang et al. (2018)

**Table 1.** MHD Turbulence Data

| Data | $B_0$ | $P_0$ | $M_A$ | $M_s$ | $\beta$ | $\langle B_A^2\rangle : \langle B_F^2\rangle : \langle B_S^2\rangle$ | $\langle B_\parallel\rangle$ | $\langle B_\perp\rangle$ | $\alpha_{1D}$ | Descriptions |
|---|---|---|---|---|---|---|---|---|---|---|
| Group1 | 0.1 | 0.05 | 1.69 | 3.11 | 0.59 | 2.16 : 0.52 : 7.32 | [0.0, 0.1] | [0.1, 0.0] | $1.76 \pm 0.02$ | super-Alfvénic and supersonic |
| Group2 | 0.1 | 2.0 | 1.72 | 0.45 | 29.21 | 2.30 : 0.01 : 7.69 | [0.0, 0.1] | [0.1, 0.0] | $2.08 \pm 0.04$ | super-Alfvénic and subsonic |
| Group3 | 1.0 | 0.05 | 0.58 | 3.16 | 0.07 | 1.92 : 7.55 : 0.53 | [0.0, 1.0] | [1.0, 0.0] | $1.56 \pm 0.03$ | sub-Alfvénic and supersonic |
| Group4 | 1.0 | 2.0 | 0.65 | 0.48 | 3.67 | 0.95 : 1.11 : 7.94 | [0.0, 1.0] | [1.0, 0.0] | $1.93 \pm 0.04$ | sub-Alfvénic and subsonic |

NOTE—The physical meanings of symbols are as follows: $B_0$ – initial magnetic field strength; $P_0$ – initial gas pressure; $M_A$ – Alfvénic Mach number; $M_s$ – sonic Mach number; $\beta$ – plasma parameter; $\langle B_i^2\rangle$ denotes the magnetic energy corresponding to the Alfvénic ($i$ = A), fast ($i$ = F), and slow ($i$ = S) modes, respectively; $\langle B_\perp\rangle$ and $\langle B_\parallel\rangle$ – mean magnetic fields strength perpendicular and parallel to the LOS, respectively; $\alpha_{1D}$ – fitted spectral index of the 3D magnetic power spectrum and its standard error. The fitting range spans from the midpoint of the weak-to-strong turbulence transition interval to the dissipation scale, which is fixed at $k = 10^{1.5}$, with k representing the wavenumber.

rotating region are spatially separated. Following R.-Y. Wang et al. (2021), we introduce scaling factors for the Faraday-rotating region relative to the emitting region: $n_e^{rot} = \alpha n_e$ and $B^{rot} = \beta B$, with $\alpha = 1$ and $\beta = 0.8$. This choice represents a foreground screen with a slightly weaker magnetic field than the emitting region while keeping the density unchanged. All other parameters remain the same as the synchrotron-emitting region.

As for differential Faraday depolarization, our synthetic observations employ the layer-by-layer line-of-sight integration scheme described in C. A. Herron et al. (2018). For the $i$-th layer ($i = 0, 1, \ldots, 512$), we compute the Faraday rotation measure $\mathrm{RM}_i$ and solve for the local Stokes parameters $Q_i$, $U_i$, and intensity $I_i$. After propagating through all slices along the sightline, we obtain the integrated total $Q$, $U$, and $I$, from which the 2D polarization degree map is constructed. For the case of wavelength-independent depolarization, we apply the same integration procedure while omitting the cumulative Faraday rotation.

As noted above, nonthermal relativistic electrons generate synchrotron emission following a power-law energy distribution (see Eq. (1)). For Faraday rotation calculations, we require the distribution of thermal electrons, which are assumed to trace the proton density field from our MHD simulations. This implies that the thermal electrons responsible for Faraday rotation share the same spatial distribution as the MHD gas.

## 4. NUMERICAL RESULTS

### 4.1. *Influence of Depolarization Processes on Polarization Degree*

#### 4.1.1. *Influence of Effective Turbulence Level*

Following the theoretical basis for the wavelength-independent depolarization processes described in Sect. 2, we employ Eqs. (5) and (10) to numerically derive the polarization degree in different turbulence regimes. With each group of the four turbulence regimes listed in Table 1, the upper panels of Fig. 2 shows the probability density distributions (PDF) of the polarization degree from independent-$\lambda$ depolarization at two limiting cases of the perpendicular LOS mean magnetic field strength, $\langle B_\perp\rangle_{\min}$ and $\langle B_\perp\rangle_{\max}$. From these panels, we observe that when $\langle B_\perp\rangle$ is near its maximum, the mean polarization degree of sub-Alfvénic turbulence is significantly higher than that of super-Alfvénic turbulence: for super-Alfvénic cases, $\langle p\rangle \approx 0.208$–$0.234$, while for sub-Alfvénic cases, $\langle p\rangle \approx 0.577$–$0.592$. This indicates stronger magnetic field ordering in sub-Alfvénic turbulence, as the PDF of $p$ for sub-Alfvénic turbulence is sharply peaked at high polarization degrees, whereas super-Alfvénic turbulence shows a broad, low-$p$ distribution. When $\langle B_\perp\rangle$ is near its minimum, the mean polarization degrees of all turbulence models decrease: for super-Alfvénic cases, $\langle p\rangle \approx 0.195$–$0.229$ (negligible change), while for sub-Alfvénic cases, $\langle p\rangle$ drops to $\approx 0.299$–$0.315$. The reduction is far more pronounced for sub-Alfvénic turbulence, suggesting its depolarization effect is more sensitive to variations in the mean magnetic field strength.

As shown in the lower panels of Fig. 2, the mean polarization degree of sub-Alfvénic turbulence is significantly lower in the differential FR case (simultaneously including both wavelength-independent and differential FR depolarization) than in the independent-$\lambda$ case (upper panels). For example, for sub-Alfvénic and supersonic turbulence, the maximum mean polarization degree decreases from 0.592 (independent-$\lambda$, at $\langle B_\perp\rangle_{\max}$) to 0.530 (differential FR, at $\langle B_\parallel\rangle_{\min}$), while super-Alfvénic turbulence shows only minor changes in mean polarization degree between the two cases (e.g., super-Alfvénic and supersonic turbulence: $\langle p\rangle = 0.208$ vs. 0.179 at $\langle B_\perp \rangle_{\max}$ for independent-$\lambda$ and $\langle B_\parallel \rangle_{\min}$ for differential FR). For the polarization degree probability density function at $\langle B_\perp \rangle_{\max}$ and $\langle B_\parallel \rangle_{\min}$, sub-Alfvénic turbulence exhibits a much broader and lower-$p$ distribution in differential FR, whereas super-Alfvénic turbulence maintains nearly identical PDF shapes between the two cases. This finding demonstrates that differential FR enhances the net depolarization effect, particularly in ordered magnetic fields. The differential phase shifts of polarized emission across different LOS depths introduced by FR drive the cancellation of orthogonal polarization components, which further reduces the observed net polarization degree and broadens the distribution of polarization degrees.

To further explore how the turbulence level (by changing the components of the mean magnetic field) modulate the depolarization process, we analyze in Fig. 3 the average polarization degree $\langle p \rangle$ as functions of $\langle B_\perp \rangle$ and $\langle B_\parallel \rangle$ for the wavelength-independent and differential FR depolarization mechanisms to quantify changes in signal coherence, respectively. For sub-Alfvénic turbulence, $\langle p \rangle$ decreases monotonically with increasing $\langle B_\parallel \rangle$, whereas super-Alfvénic turbulence shows very weak sensitivity to variations in the LOS mean magnetic field strength, with $\langle p \rangle$ remaining nearly constant across the full range of $\langle B_\parallel \rangle$.

These results further demonstrate that differential FR significantly enhances the depolarization effect compared to the wavelength-independent-only case. Specifically, for sub-Alfvénic turbulence, the mean polarization degree $\langle p \rangle$ under differential FR is substantially lower than under independent-$\lambda$ at nearly all values of $\langle B_\parallel \rangle$ and $\langle B_\perp \rangle$, and its decline with increasing $\langle B_\parallel \rangle$ is also more pronounced. In contrast, super-Alfvénic turbulence exhibits negligible differences in $\langle p \rangle$ between the two mechanisms, consistent with its already low and stable polarization degree. Physically, the differential phase shifts of polarized emission introduced by Faraday depolarization along the LOS drive the cancellation of orthogonal polarization components, which further reduces the net polarized signal, broadens the probability density distribution of the polarization degree, and increases the heterogeneity of the polarization response with respect to $\langle B_\parallel \rangle$.

#### 4.1.2. *Influence of the Radiation Frequency*

Treating wavelength-independent depolarization as the intrinsic attenuation of polarization signals in turbulent media, we compare the contributions and dominant characteristics of the differential FR and beam depolarization processes. Numerical results presented in Fig. 4 reveal clear differences in the dominant depolarization mechanisms between turbulence regimes. For sub-Alfvénic turbulence (curves c, d, g and h), depolarization is dominated by the differential FR mechanism: the blue curves (differential FR-included cases) show systematically lower polarization degrees than the corresponding red curves (beam depolarization) across the entire frequency range, with the effect most pronounced at low frequencies. In contrast, for super-Alfvénic turbulence (curves a, b, e and f), depolarization is dominated by beam depolarization at high frequencies. At low frequencies, the beam depolarization saturates, and differential Faraday depolarization takes over as the dominant mechanism, leading to a significant decrease in the polarization degree.

From the perspective of spatial integration, the differing contributions of differential FR and beam depolarization arise from their distinct dependencies on path-length integration, which are governed by the dynamical properties of the turbulence. For sub-Alfvénic turbulence, the strong magnetic field effect leads to large-scale spatial coherence in the $\mathrm{RM}(\mathbf{x})$. This coherence significantly amplifies the fluctuation amplitude of the local Faraday phase difference $\Delta\chi$, driving strong randomization of the polarization plane orientation. This effect directly and strongly attenuates the polarization degree, making differential FR the dominant depolarization mechanism. In contrast, beam depolarization is a spatially integrated process, including another Faraday depolarization in the foreground region. Here, small-scale fluctuations in polarization intensity $e(\mathbf{x})$ and polarization angle $\theta(\mathbf{x})$ are strongly suppressed by the ordered magnetic field. This suppression enhances spatial correlations within the integration kernel, preventing full randomization of the phase difference, reducing the integration's fluctuation amplitude, and thus weakening the overall contribution of beam depolarization.

Correspondingly, super-Alfvénic turbulence exhibits reversed behavior in terms of mechanism dominance. Strong fluid shear enhances small-scale fluctuations in $e(\mathbf{x})$ and $\theta(\mathbf{x})$, breaking spatial correlations in the beam depolarization integral. This drives complete randomization of the polarization angle phase difference, making spatial integration highly efficient at attenuating the net polarization and establishing beam depolarization as the dominant mechanism at high frequencies; however, this effect saturates at low frequencies. In super-Alfvénic turbulence, fluid convection leads to a small-scale random distribution of $\mathrm{RM}(\mathbf{x})$, and the fluctuation amplitude of the local $\Delta\chi$ is rapidly canceled out by small-scale structures. This suppresses the randomization of the polarization plane rotation, greatly reducing the attenuation effect on the polarization degree and leaving the contribution of differential FR weak at high frequencies, where the depolarization effect is temporarily suppressed. However, at low frequencies, the strong $\lambda^2$ dependence amplifies differential FR, making it the dominant mechanism.

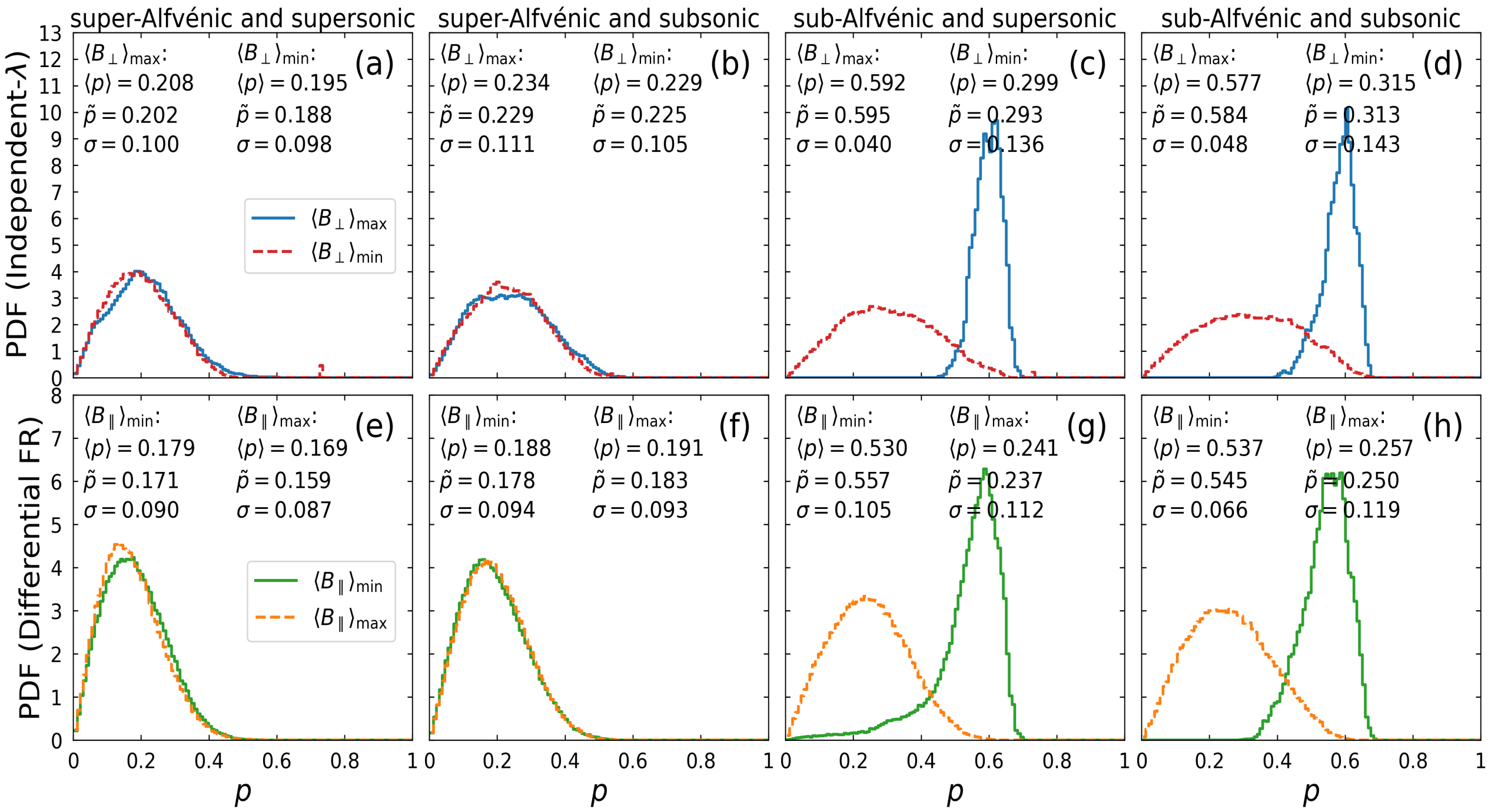

**Figure 2.** Probability density function (PDF) of the polarization degree of synchrotron radiation, corresponding to two depolarization mechanisms: wavelength-independent depolarization (upper panels) and differential Faraday depolarization (lower panels) at a low frequency of 60 MHz. Results are displayed for two mean magnetic field strengths, i.e., $\langle B_\parallel \rangle$ and $\langle B_\perp \rangle$, covering various sub-Alfvénic and super-Alfvénic turbulence cases, where $\langle B_\parallel \rangle_{\rm min}$ and $\langle B_\parallel \rangle_{\rm max}$ correspond to differential Faraday rotation, while $\langle B_\perp \rangle_{\rm min}$ and $\langle B_\perp \rangle_{\rm max}$ correspond to independent-$\lambda$ depolarization. In each panel, $\langle p \rangle$, $\tilde{p}$, and $\sigma$ denote the mean, median, and standard deviation of the polarization degree distribution, respectively. It should be noted that differential FR incorporates both wavelength-independent depolarization and differential FR-induced depolarization.

## 4.2. *Influence of Depolarization Processes on Power Spectrum of Polarization Intensity*

### 4.2.1. *Influence of Effective Turbulence Level*

We investigate the influence of turbulence level characterized by varying the mean magnetic field strength on the power spectra across different turbulence regimes. Fig. 5 shows that sub-Alfvénic and super-Alfvénic turbulence display divergent spectral behaviors with varying ordered magnetic field components, with $\langle B_\perp \rangle$ for the independent-$\lambda$ depolarization and $\langle B_\parallel \rangle$ for the differential FR depolarization. The vertical shaded band in the figure marks the weak-to-strong turbulence transition.[5] The inertial-range scaling of the two-dimensional polarization intensity power spectrum undergoes a distinct transition (see panels (c)–(d) of Fig. 5) in sub-Alfvénic cases, where the spectral slope varies with increasing $\langle B_\perp \rangle$ for independent-$\lambda$. This transition does not originate from the intrinsic properties of the turbulent energy cascade, but is governed instead by the depolarization effect modulated by the mean magnetic field. At large values of $\langle B_\parallel \rangle$, the enhanced suppression of power at small scales highlights the field-dependent nature of the differential FR depolarization process. Furthermore, at a frequency of 60 MHz, the independent-$\lambda$ and differential FR depolarization power spectra exhibit significant differences under super-Alfvénic and sub-Alfvénic turbulent conditions.

In contrast, the power-law scaling of the super-Alfvénic turbulence spectrum remains largely unchanged as $\langle B_\parallel \rangle$ and $\langle B_\perp \rangle$ vary (panels (a)–(b)). This spectral stability arises directly from the intrinsic properties of super-Alfvénic turbulence, where the turbulent kinetic energy dominates over magnetic energy. The weak mean magnetic field in this regime exerts only a minor influence on the turbulent cascade, so the ratio of fluctuating to mean field strength has little impact on the overall spectral shape.

[5] The wavenumber range of transition region follows from the turbulence partition, which generalizes the GS95 to both sub- and super-Alfvénic regimes (A. Lazarian & E. T. Vishniac 1999; A. Lazarian 2006). Using $k \propto 1/l$ and the injection wavenumber interval $k_{\rm inj} = (2.0, 3.0)$, the transition scales $l_{\rm sup} = L_{\rm inj} M_{\rm A}^{-3}$ for super-Alfvénic turbulence and $l_{\rm sub} = L_{\rm inj} M_{\rm A}^2$ for sub-Alfvénic turbulence correspond to $k_{\rm trans} = k_{\rm inj} M_{\rm A}^3$ and $k_{\rm trans} = k_{\rm inj}/M_{\rm A}^2$, respectively.

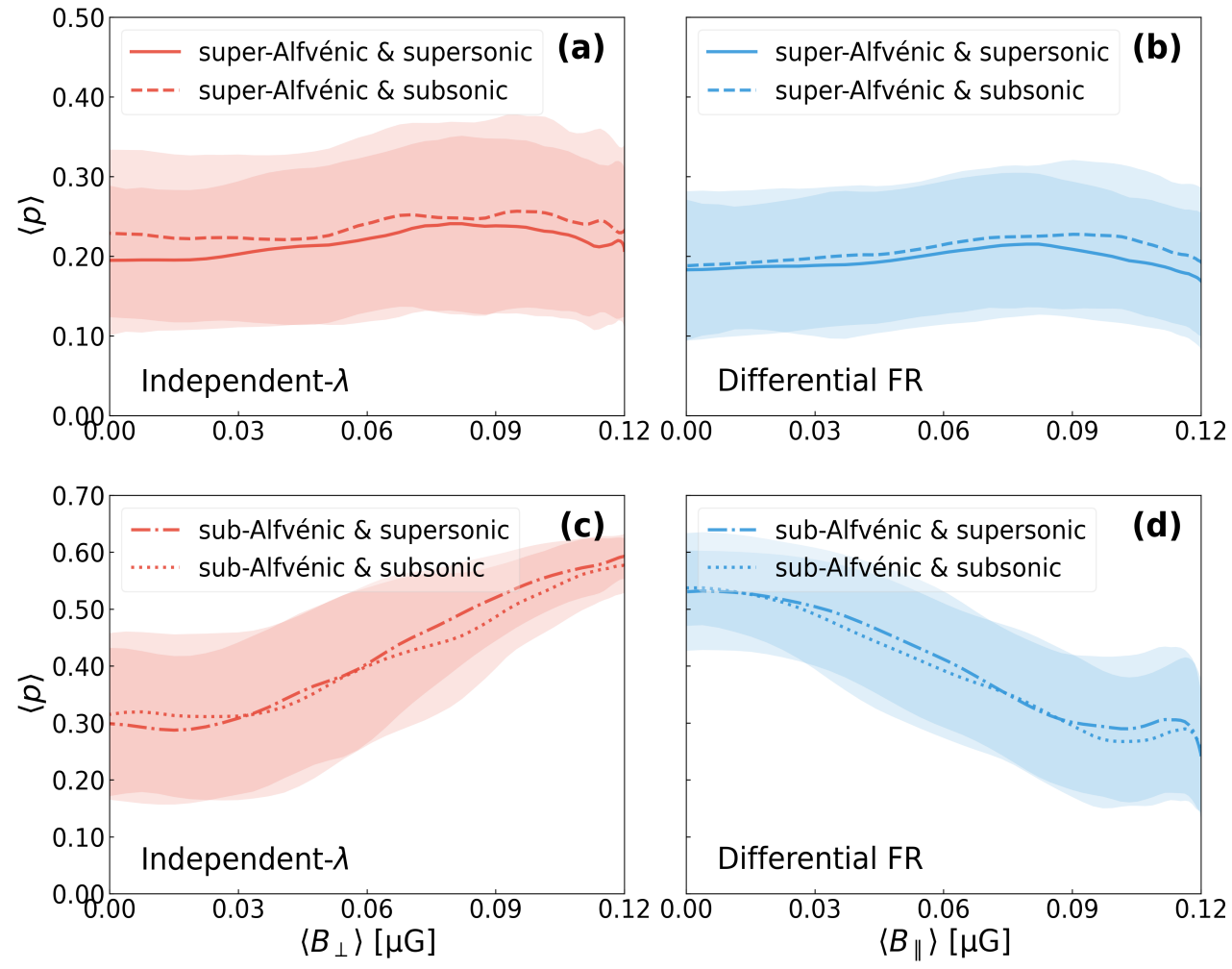

**Figure 3.** Average polarization degree at a frequency of 60 MHz as functions of the mean magnetic field components, i.e., $\langle B_{\perp}\rangle$ and $\langle B_{\parallel}\rangle$, for various turbulence regimes. The left and right columns correspond to independent-$\lambda$ and differential Faraday depolarization, respectively. The top and bottom rows correspond to super-Alfvénic (regions a, b) and sub-Alfvénic (regions c, d) turbulence regimes, respectively. The shaded bands represent the 1 $\sigma$ standard deviation of the polarization degree over all pixels at each fixed magnetic field orientation.

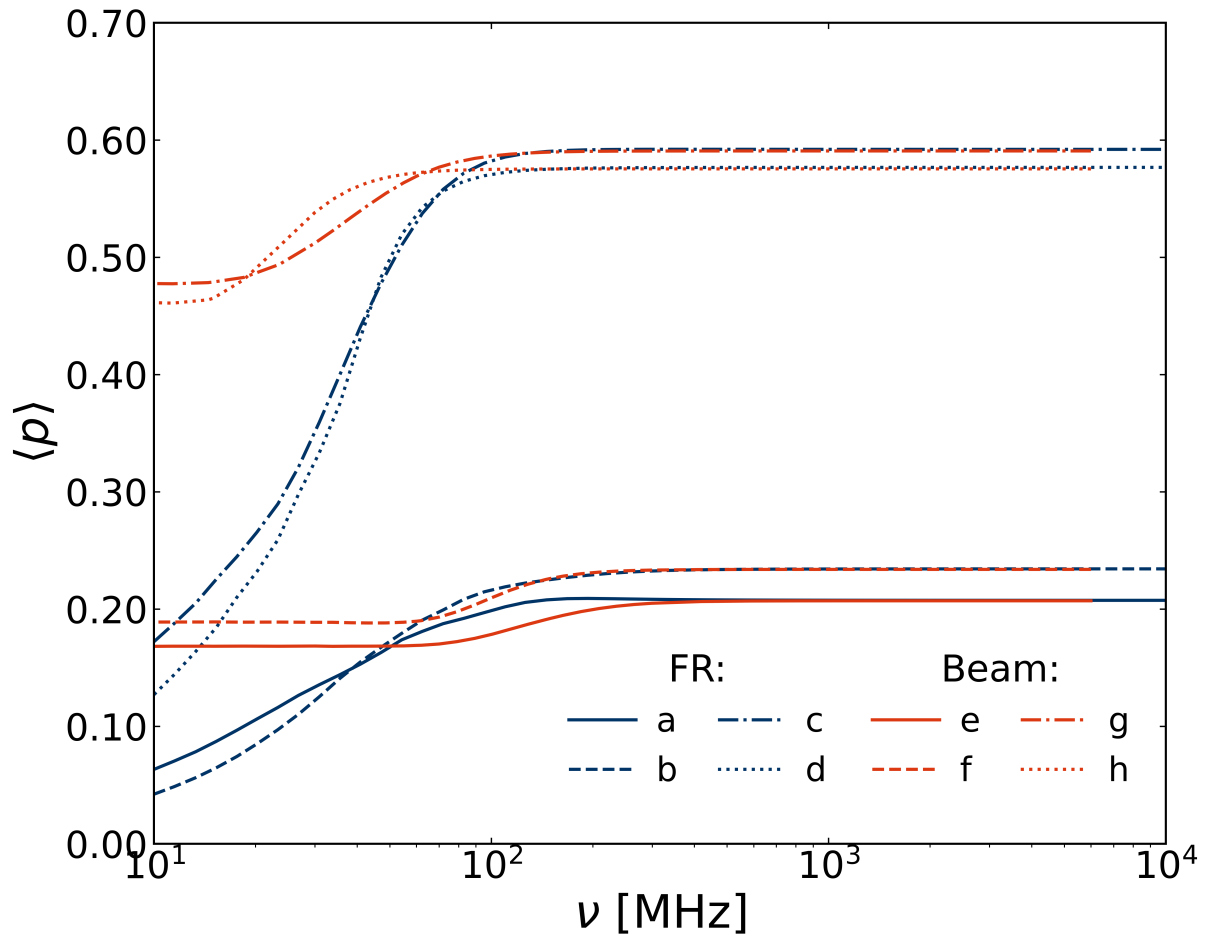

**Figure 4.** The variation of the average polarization degree $\langle p\rangle$ with observational frequency, for a zero LOS mean magnetic field and a beam FWHM of approximately 2 pc (1 pixel). Red curves represent cases including beam depolarization, while blue curves correspond to cases incorporating differential FR depolarization. The line styles denote different turbulence regimes: solid (super-Alfvénic, supersonic), dashed (super-Alfvénic, subsonic), dot-dashed (sub-Alfvénic, supersonic), and dotted (sub-Alfvénic, subsonic).

#### 4.2.2. *Influence of the Radiation Frequency*

In turbulent magnetized media, stochastic fluctuations in magnetic field orientation render wavelength-independent depolarization an unavoidable fundamental physical process and a necessary prerequisite for analyzing other depolarization mechanisms, including differential FR depolarization and beam depolarization. To investigate how these two depolarization mechanisms depend on observational frequency, we compute the power spectra of synchrotron polarization intensity across various turbulence regimes and interested frequency ranges (as shown in Fig. 4) in Figs. 6 and 7.

For differential Faraday depolarization, the inertial-range scaling of the two-dimensional polarization intensity power spectrum, and its evolution with $\langle B_{\parallel}\rangle$ (see also Fig. 5), are statistically consistent with the predictions of wavelength-independent depolarization alone at sufficiently high frequencies. Including differential FR in the composite depolarization model introduces a clear frequency dependence in the polarization intensity power spectrum, as illustrated in Fig. 6. To isolate the effect of differential FR, we focus on a low-frequency range from 10 to $10^4$ MHz, where Faraday effects are strong and easily distinguishable. As shown in the panels, decreasing frequency $\nu$ significantly alters the spectral distribution: power in the large-$k$ (small-scale) region is suppressed by enhanced FR depolarization, while power in the small-$k$ (large-scale) region is amplified by FR-induced fluctuations. As a result, the large-$k$ spectral index decreases with lower frequency, indicating that small-scale polarization structure becomes dominated by a progressively shallower spectrum. This evolution departs from the pure inertial-range power-law scaling and matches the theoretical predictions of LP16 for polarized radiation transfer in magnetized turbulent media.

When including the beam depolarization effect, we focus our analysis on the intermediate-to-high frequency range, where differences between wavelength-independent depolarization (from the emitting source region) and differential Faraday depolarization from the foreground are relatively mild. The polarization intensity power spectra derived from composite depolarization statistics for four distinct turbulence regimes are presented in Fig. 7.

As the observing frequency changes, only sub-Alfvénic turbulence exhibits significant deviations from the theoretical inertial-range power-law scaling at the

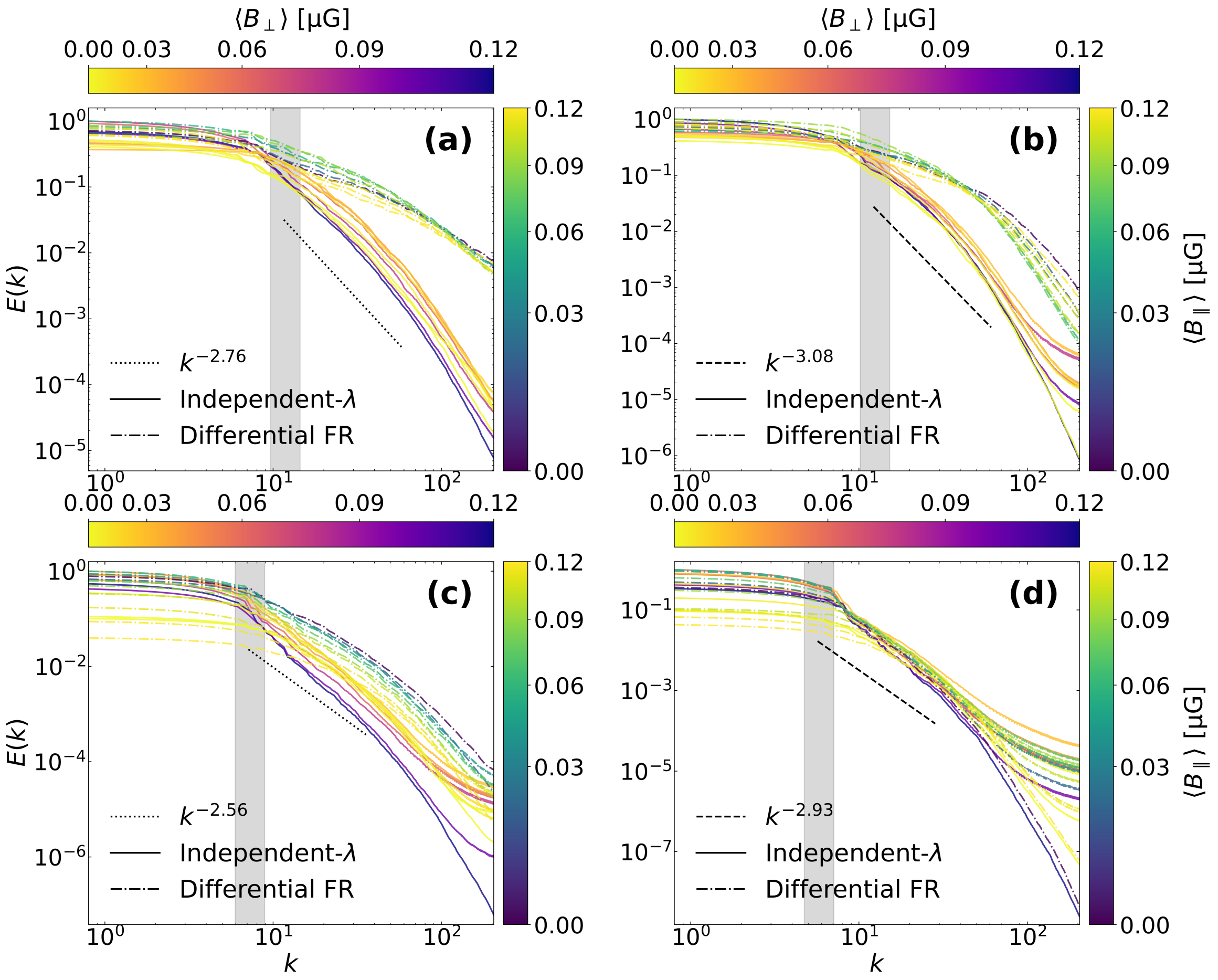

**Figure 5.** Power spectra of synchrotron polarization intensity incorporating both independent-$\lambda$ and differential FR depolarization at 60 MHz. The vertical black bands denote the transition from weak to strong turbulence. Panels (a)–(d) correspond to the super-Alfvénic and supersonic, super-Alfvénic and subsonic, sub-Alfvénic and supersonic, and sub-Alfvénic and subsonic cases, respectively. The labeled color bar indicates variations in the mean magnetic fields, with $\langle B_\perp \rangle$ for independent-$\lambda$ and $\langle B_\parallel \rangle$ for differential FR.

large-$k$ region, whereas super-Alfvénic turbulence remains nearly unaffected in the inertial range. This behavior arises from the amplification of the RM contribution in the wavelength-dependent phase factor $\exp\left[2i\mathrm{RM}(\mathbf{x})\lambda^2 - 2i\mathrm{RM}(\mathbf{x}+\mathbf{s})\lambda^2\right]$ as $\lambda$ increases. Small-scale MHD turbulence exhibits local spatial coherence, such that the increase in $\lambda^2$ fails to fully randomize its polarization phase. In contrast, large-scale structures are more easily depolarized by this wavelength-dependent effect. Consequently, the power spectrum shows a relative enhancement at large $k$ (small scales) and a suppression at small $k$ (large scales), leading to the observed flattening of the spectrum and a clear deviation from the intrinsic turbulence power-law scaling.

Furthermore, the degree of deviation depends strongly on the turbulence regime. For super-Alfvénic turbulence (panels (a) and (b)), the spectra remain a relatively small dispersion with only a weak deviation from the inertial range scaling, even at the largest wavelengths considered. In contrast, sub-Alfvénic turbulence (panels (c) and (d)) exhibits far more drastic spectral evolution: the spectra are significantly more dispersed, with a pronounced and systematic deviation from the theoretical scaling. This key distinction can be directly explained by the relationship of $\delta B_\parallel = (\delta B_{\rm rms}/\langle B \rangle) \cdot \langle B \rangle$, where $\delta B_{\rm rms}$ is the root-mean-square total magnetic field fluctuation, $\delta B_{\rm rms}/\langle B \rangle$ the relative fluctuation level, and $\langle B \rangle$ the mean background magnetic field strength. In super-Alfvénic turbulence, the mean field $\langle B \rangle$ is intrinsically weak. Even when the relative fluctuation level $\delta B_{\rm rms}/\langle B \rangle$ approaches unity, the resulting absolute parallel fluctuation $\delta B_\parallel$ remains small, producing only low-amplitude small-scale RM differences ($\Delta$RM). These difference in the amplitude of $\Delta$RM leads to a different degree of variation in the frequency-dependent phase difference $\lambda^2\Delta$RM, and thus to a more pronounced

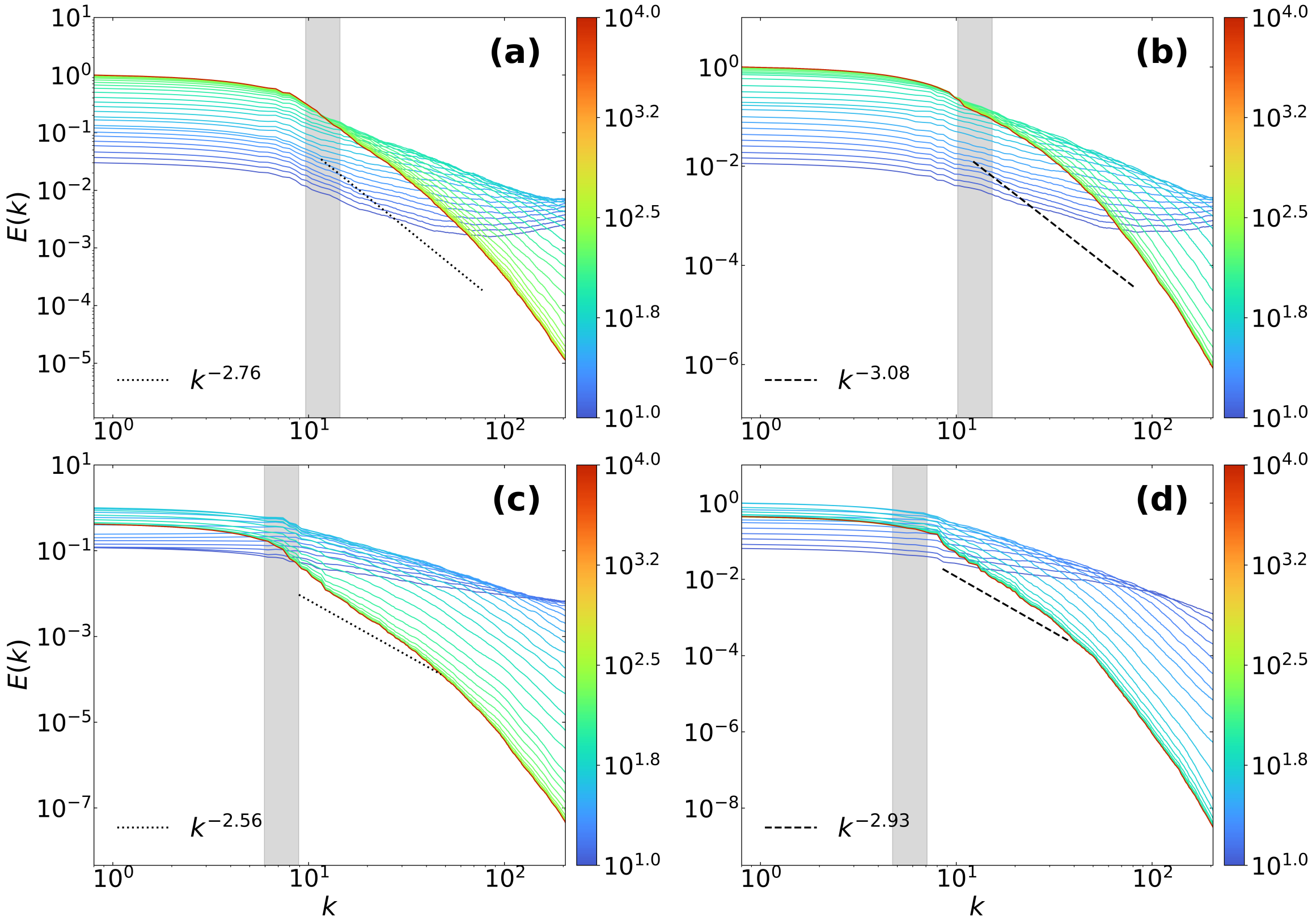

**Figure 6.** Power spectra of synchrotron polarization intensity for various turbulence regimes and frequencies, incorporating both wavelength-independent depolarization and differential FR depolarization. The light purple vertical bands indicate the weak-to-strong turbulence transition. Panels (a)–(d) represent the super-Alfvénic and supersonic, super-Alfvénic and subsonic, sub-Alfvénic and supersonic, and sub-Alfvénic and subsonic cases, respectively. The color bar shows frequency variation in MHz, with zero mean magnetic field strength along the LOS direction.

frequency-induced deviation of the polarization power spectrum in sub-Alfvénic turbulence.

#### 4.2.3. *Influence of Beamwidth*

Based on the statistical average formula for polarization intensity given in Eq. (12), we further explore beam-induced spectral variations at a fixed observing frequency, by varying the beam width parameter $t$ (defined in Eq. (15)). This parameter, which directly relates to the beam Full Width at Half Maximum (FWHM) $\omega$, acts as a key factor shaping the power-law scaling of the depolarized power spectrum by modulating the spatial contribution weights.

We assess the impact of beam width on the recovered spectral features under different turbulent environments by computing polarization intensity power spectra for four representative turbulence regimes, as shown in Fig. 8. Across all regimes, the spectra consistently show that as the beam width increases, deviations from the inertial-range power law become more pronounced. Critically, the character and rate of this evolution differ dramatically between turbulence regimes: the spectral deviation proceeds far more rapidly for super-Alfvénic turbulence than for sub-Alfvénic turbulence, with qualitatively distinct behaviors in each case.

For super-Alfvénic turbulence, the spectrum is nearly unperturbed when the beam width is small, with the power law closely following the theoretical inertial-range scaling. As the beam broadens, however, the spectrum systematically shifts downward at large $k$, leading to a progressive increase in the degree of deviation from the ideal scaling. This behavior arises because the RM/polarization field has weaker coherence in super-Alfvénic turbulence. Spatial averaging over a wider beam rapidly disrupts this weak coherence, leading to quick randomization of the phase factor $\exp[2i\lambda^2\Delta\mathrm{RM}]$, which significantly perturbs the inertial-range power-law scaling.

In contrast, sub-Alfvénic turbulence exhibits qualitatively different spectral behavior even at small beam widths: the spectrum already shows a clear upward deviation from the theoretical inertial-range power-law scaling at intermediate-to-large $k$ values. As the beam width increases, the spectrum exhibits a more complex deviation pattern from the power-law scaling. This distinct behavior can be attributed to the stronger coherence of the RM/polarization field in sub-Alfvénic turbulence. Under these conditions, beam broadening is far less effective at fully randomizing the correlated fluctuation characteristics of the turbulence, resulting in a slower, more gradual onset of spectral deviation compared to the super-Alfvénic case.

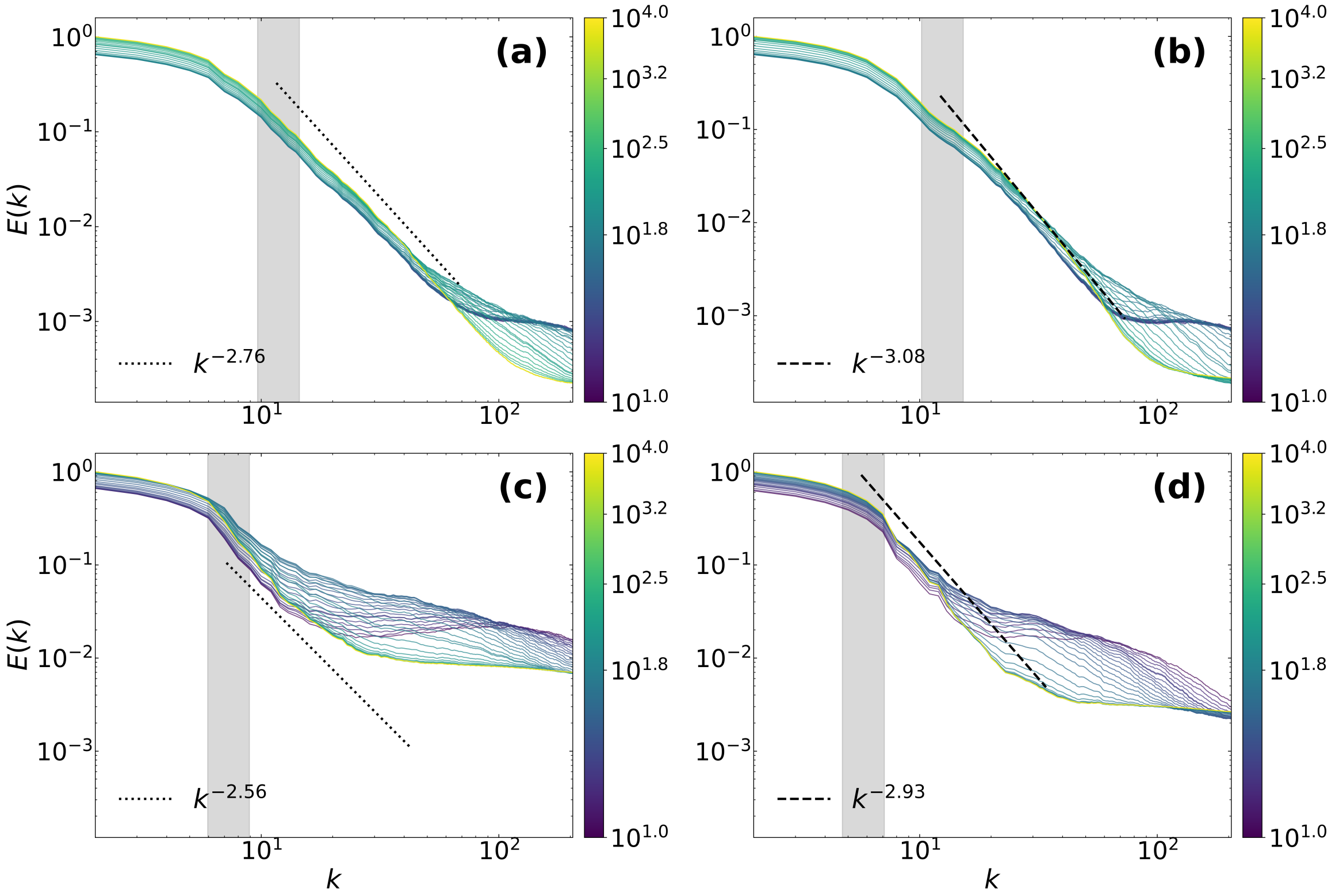

**Figure 7.** Polarization intensity power spectra incorporating both wavelength-independent depolarization (from emitting source region without FR depolarization) and beam one (including Faraday depolarization from foreground region). Panels (a)–(d) show the spectra derived from the composite depolarization statistics, corresponding to the super-Alfvénic and supersonic, super-Alfvénic and subsonic, sub-Alfvénic and supersonic, and sub-Alfvénic and subsonic cases, respectively, with a beam full width at half maximum (FWHM) of approximately 2 pc (1 pixel). The colorbar indicates the observational frequency variation in MHz. The vertical shaded areas indicate the range of the weak-to-strong turbulence transition.

## 5. DISCUSSION

Taking the ISM as an isothermal system, this work employed numerical experiments to investigate depolarization processes, while neglecting the dynamic evolution of the ISM driven by heating and radiative cooling. Additionally, the model adopts a single-phase approximation for the ISM, overlooking its complex multiphase structure of coexisting cold, warm, and hot phases that contribute differently to polarization and FR processes (A. Bracco et al. 2022). Furthermore, bandwidth depolarization, induced by the finite non-zero bandwidth of radio frequency channels, can affect polarization measurements. Nevertheless, most observational setups are optimized to reduce this effect, making it negligible for the majority of analysis. For a representative LOFAR channel at 60 MHz ($\Delta\nu \simeq 3 \times 10^{-3}$ MHz), our estimated bandwidth depolarization is well below 1% and can be safely neglected (M. A. Fine et al. 2023). While our fiducial model adopts a 1 kpc integration path, increasing it to 10 kpc would increase the Faraday depth ($\phi_{\rm FR} \propto n_{\rm e} B_{\parallel} L$), thereby enhancing depolarization and allowing significant depolarization to occur at higher observing frequencies. Another effect not considered in the present study is synchrotron self-absorption (SSA), which can become important in optically thick emitting regions. N.-N. Guo et al. (2026) showed that SSA significantly suppresses polarized intensity and modifies its spatial anisotropy and power spectrum, while preserving the underlying turbulence scaling in the inertial range. Incorporating SSA into our framework would enable a more comprehensive assessment of depolarization effects in optically thick astrophysical environments.

Wavelength-independent depolarization dominates the attenuation of polarization at short wavelengths, as evidenced by 3.5 cm observations of the M51 halo (C. Shneider et al. 2014), where wavelength-dependent effects are strongly suppressed; the Coma Cluster radio halo, meanwhile, serves as a theoretically purer candidate for this effect (B. J. Burn 1966). Differential FR depolarization prevails in sub-Alfvénic turbulence, which is typically found in dense, strongly magnetized environments such as galactic disks (C. L. Van Eck et al. 2015). By contrast, beam depolarization is the primary mechanism in flow-dominated super-Alfvénic turbulence, which commonly occurs in diffuse, low-density environments including Galactic halo and cluster radio halo. We can incorporate this effect in Cygnus A and analogous sources, where the majority of observed Faraday depolarization originates from foreground material rather than the source itself (G. Brunetti & T. W. Jones 2014; C. L. Carilli et al. 1991).

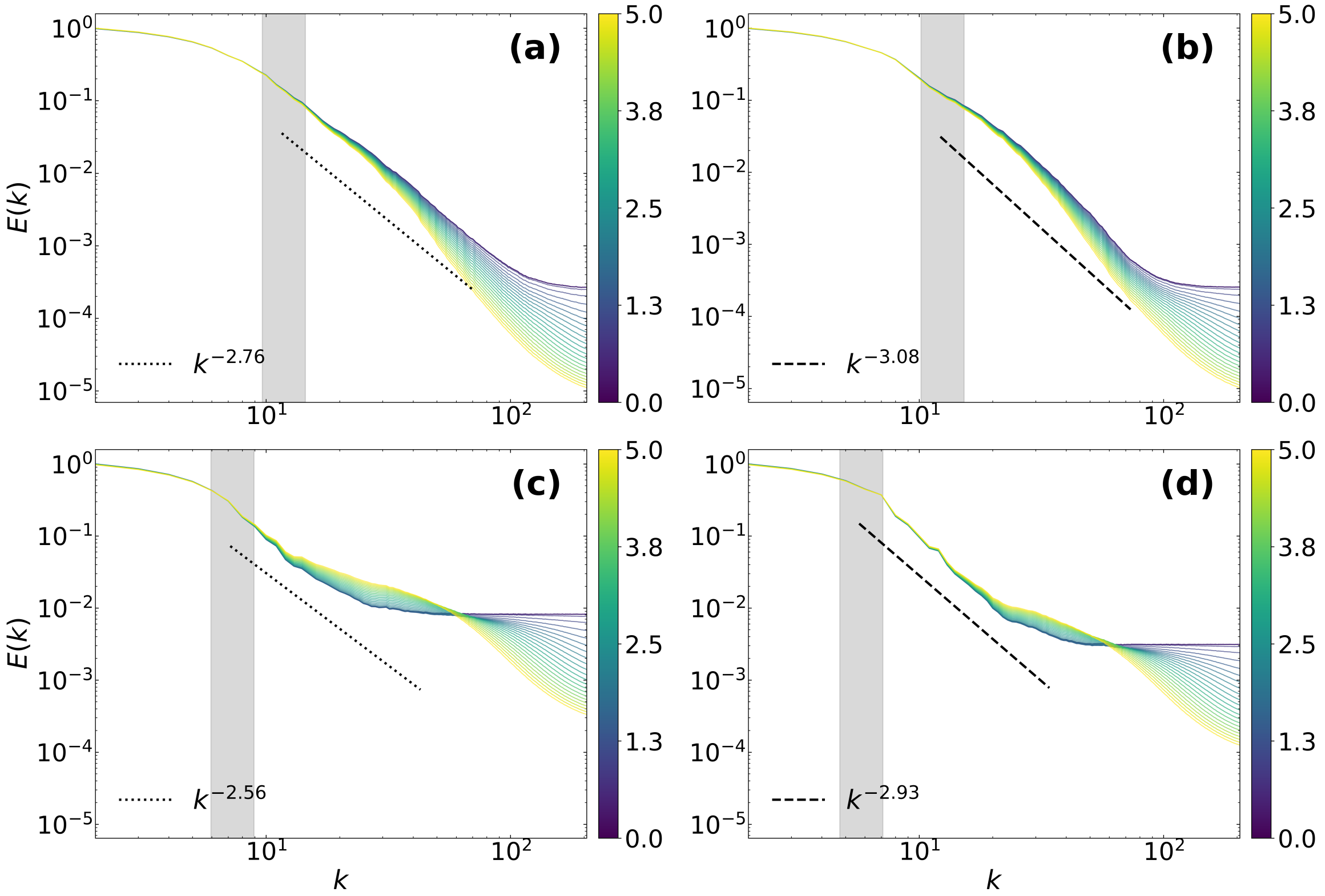

**Figure 8.** Power spectra of synchrotron polarization intensity for different beamwidths and turbulence regimes at a fixed observational frequency of $10^4$ MHz. Panels (a)–(d) show the spectra derived from the composite depolarization statistics, corresponding to the super-Alfvénic and supersonic, super-Alfvénic and subsonic, sub-Alfvénic and supersonic, and sub-Alfvénic and subsonic cases, respectively. The color bar shows that the beam FWHM ranges from 0.1 to 5 pixels, corresponding to spatial scales ranging from 0.2 to 9.8 pc. The vertical shaded areas indicate the range of the weak-to-strong turbulence transition.

When exploring beam depolarization, we considered different foreground and background parameters by via a scaling factor to simplify calculations. This simplified scheme is valid under ideal special scenarios but fails to fully encompass realistic astrophysical environments. As the background, the Galactic halo exhibits different polarized radiation properties compared to the foreground Galactic disk, which is characterized by complex radiation transfer effects (E. Carretti et al. 2010). Additionally, extragalactic background sources display significant physical discrepancies relative to the foreground Milky Way (S. P. O'Sullivan et al. 2012). Our current simulations do not incorporate such environmental differences; future work can further investigate their specific impacts. Furthermore, we note that the one-pixel beam adopted in this work to investigate the effect of beam width on the power spectrum is slightly undersampled and is intended solely for illustrative purposes.

We anticipate that our synthetic analysis framework is applicable to real observations obtained with the VLA, LOFAR, and FAST. For the fixed 2 pc beam adopted in this work, the angular–physical scale conversion gives $\theta_{\rm arcsec} = 412.53/D_{\rm kpc}$. Using this relation, the VLA is capable of probing sources across kpc to Mpc distances, whereas FAST is mostly suited to Galactic targets within $\sim 2$ kpc. Our adopted frequencies, namely 60 MHz and the range 10–$10^4$ MHz, overlap substantially with the observing bands of these telescopes, particularly LOFAR. We note that low-frequency observations require narrow channelization or RM-synthesis to account for bandwidth depolarization (M. A. Fine et al. 2023). However, our calculations are monochromatic, evaluated at discrete central frequencies, and therefore omit continuous bandwidth-integration effects. A robust application of our method to real data demands sufficient angular resolution to resolve small-scale turbulent structures, adequate polarization sensitivity to detect the subtle statistical signatures of MHD turbulence, and careful mitigation of instrumental noise and beam-averaging effects.

Unlike real observations, our simulations allow us to independently control differential Faraday rotation, thereby isolating its imprint on polarization statistics. In real radio data, however, this effect is heavily intertwined with other depolarization processes, noise, and beam convolution (A. H. Waelkens et al. 2009), making it difficult to directly disentangle. Nevertheless, our synthetic results serve as theoretical benchmarks; by comparing observational data with synthetic models, the dominant depolarization mechanism can be inferred.

Our study mainly focuses on the influence of depolarization processes on polarization degree and power spectra of synchrotron polarization fluctuations. Depolarization effects may hinder the accurate diagnosis of typical turbulence features such as anisotropy, intermittency and magnetization strength, which require further in-depth research. Furthermore, comprehensive exploration of various depolarization mechanisms helps constrain the three-dimensional magnetic field structure of the Milky Way (J. L. Han et al. 2004).

This work presents a general framework for identifying the dominant depolarization mechanisms in various turbulence regimes, which provides a quantitative reference for the study of astrophysical turbulence and magnetic fields using radio polarization observations. Benefiting from the development of next-generation radio facilities such as the Square Kilometre Array, high-quality observational data with unprecedented wide-band coverage, angular resolution, and polarization sensitivity will be obtained. Our results can be applied to understand specific astrophysical objects (e.g., nearby spiral galaxies, galaxy clusters, and radio galaxies) and offer valuable references for relevant studies of cosmic microwave background polarization, serving as useful support for future multi-scale investigations of magnetic fields and turbulence in radio astronomy.

## 6. SUMMARY

We numerically studied three typical depolarization processes in magnetized turbulent media, including wavelength-independent, differential FR, and beam depolarization, and analyzed their effects on polarization degree and power spectrum under different turbulence states, observational frequencies and telescope beam sizes. Our main numerical findings are summarized as follows:

1. Sub-Alfvénic turbulence exhibits significantly higher polarization degree compared to super-Alfvénic turbulence, indicating stronger magnetic field ordering in sub-Alfvénic turbulence; however, sub-Alfvénic turbulence is also more sensitive to variations in turbulence level, with its mean polarization degree decreasing by about 50% when the field weakens.

2. Differential FR significantly enhances the depolarization effect, particularly in sub-Alfvénic turbulence, where the mean polarization degree decreases from $\simeq 0.6$ (independent-$\lambda$ case) to $\simeq 0.5$ (differential FR case), while super-Alfvénic turbulence shows negligible differences between the two mechanisms.

3. Dominant depolarization mechanism differs between turbulence regimes: for sub-Alfvénic turbulence, differential FR dominates due to large-scale spatial coherence in $\mathrm{RM}(x)$; for super-Alfvénic turbulence, beam depolarization dominates at high frequencies because strong fluid shear enhances small-scale fluctuations that break spatial correlations in the integration kernel. At low frequencies, the beam depolarization saturates, and differential Faraday depolarization takes over as the dominant mechanism.

4. Sub-Alfvénic turbulence exhibits spectral transitions with increasing $\langle B_\perp \rangle$ for independent-$\lambda$, where the spectral slope varies due to depolarization effects, while super-Alfvénic turbulence maintains stable power-law scaling because turbulent kinetic energy dominates over magnetic energy.

5. Differential Faraday depolarization significantly alters the polarization power spectrum by suppressing power at large $k$ (small scales) and amplifying it at small $k$ (large scales), causing spectral indices that depart from the theoretical inertial-range scaling.

6. Super-Alfvénic turbulence shows rapid spectral deviation from the theoretical inertial-range power-law scaling as beam width increases due to weak parallel magnetic field fluctuations, whereas sub-Alfvénic turbulence exhibits a more complex deviation from the power-law scaling.

These results clarify depolarization characteristics across different turbulence regimes. They provide the-oretical foundations for diagnosing magnetic turbulence using both polarization degree and power spectra.

## ACKNOWLEDGMENTS

We thank the anonymous referee for valuable comments that have greatly improved the manuscript. We are grateful to Xuewen Zhang for assistance with numerical tests. We acknowledge support from the National SKA Program of China (Grant No. 2025SKA0140100) and the National Natural Science Foundation of China (Grant Nos. 12473046, 12233002 and 12473042).

## APPENDIX

## A. 3D VECTOR ROTATION OF THE MAGNETIC FIELD

To investigate the effect of different depolarization processes on the polarization degree, we quantify the effective turbulence level by independently varying the mean magnetic field strengths along the parallel and perpendicular LOS directions. To rotate the mean magnetic field orientation relative to the line of sight, we use the total rotation matrix $\hat{F} = \hat{F}_x \hat{F}_y \hat{F}_z$ (X.-W. Zhang et al. 2023), the components of which are

$$\hat{F}_x = \begin{bmatrix} 1 & 0 & 0 \\ 0 & \cos(\varphi_x) & \sin(\varphi_x) \\ 0 & -\sin(\varphi_x) & \cos(\varphi_x) \end{bmatrix}, \tag{A1}$$

$$\hat{F}_y = \begin{bmatrix} \cos(\varphi_y) & 0 & -\sin(\varphi_y) \\ 0 & 1 & 0 \\ \sin(\varphi_y) & 0 & \cos(\varphi_y) \end{bmatrix}, \tag{A2}$$

$$\hat{F}_z = \begin{bmatrix} \cos(\varphi_z) & \sin(\varphi_z) & 0 \\ -\sin(\varphi_z) & \cos(\varphi_z) & 0 \\ 0 & 0 & 1 \end{bmatrix}, \tag{A3}$$

where $\varphi_{m=x,y,z}$ are the rotation angles about the $x$, $y$, and $z$ axes, respectively. In this work, we only rotate the data cube around the $y$-axis. For data cube $\mathcal{R}(\mathbf{r})$, We obtain rotated data cubes $\hat{F}\mathcal{R}(\hat{F}^{-1}\mathbf{r})$ from this 3D vector rotation, where $\boldsymbol{r}$ denotes the position vector. This procedure simultaneously changes both the LOS and plane-of-sky magnetic field components, effectively varying $\langle B_\perp \rangle$ and $\langle B_\parallel \rangle$ between their maximum values and zero. It is important to note that $\langle B_\parallel \rangle$ and $\langle B_\perp \rangle$ are not independent, as the total mean field is fixed; thus, these trends reflect variations along a single rotation angle. Furthermore, rotating the cube cannot genuinely mimic independent simulations with different initial $B_0$. While rotation preserves the full intrinsic 3D turbulence statistics (e.g., $\delta B$, $M_{\rm A}$, $M_{\rm S}$), it only modifies relative quantities such as the Faraday depth $\phi_{\rm FR}$ and polarization intensity $P$. Our intention is simply to isolate how variations in the LOS magnetic component shape the depolarization mechanism, rather than to substitute for distinct independent simulations.

## B. STRUCTURE FUNCTION ANALYSIS OF POLARIZATION INTENSITY

To verify the robustness of the spectral properties obtained from the power spectrum analysis, we perform a structure function analysis of the polarization intensity, using differential Faraday rotation depolarization as an illustrative example. The second-order structure function of the complex polarization intensity $P$ is defined as ${\rm SF}_2(r) = \langle [P(\boldsymbol{x} + \boldsymbol{r}) - P(\boldsymbol{x})]^2 \rangle$. According to the theoretical relations (A. S. Monin et al. 1976), a 2D power spectrum of $E(k) \propto k^{-\alpha_{\rm 2D}}$ corresponds to the structure function scaling of ${\rm SF}_2(r) \propto r^{\alpha_{\rm 2D}-1}$. For homogeneous MHD turbulence, the line-of-sight projection relates the fitted 3D spectral index $\alpha_{\rm 1D}$ to the 2D index via $\alpha_{\rm 2D} = \alpha_{\rm 1D} + 1$. Therefore, the expected 2D structure function slope is $\beta_{\rm 2D} = \alpha_{\rm 2D} - 1 = \alpha_{\rm 1D}$. Canonically, this translates to $r^{5/3}$ for the Kolmogorov case and $r^2$ for the steeper subsonic case.

Fig. 9 shows the structure functions ${\rm SF}_2(r)$ at $\nu = 10^4$ MHz for the four turbulence regimes. The black dashed lines indicate the reference slopes, i.e., $r^{5/3}$ for the supersonic cases (panels a and c) and $r^2$ for the subsonic cases (panels b and d). As shown in the figure, the structure functions exhibit a robust power-law scaling consistent with these theoretical relations. Note that a robust power-law scaling of the structure function requires high spatial resolution to resolve a sufficiently broad inertial range; insufficient resolution tends to suppress the true power-law signature (H. Lee et al. 2016). Overall, the structure function analysis effectively supports the conclusions drawn from the power spectrum analysis.

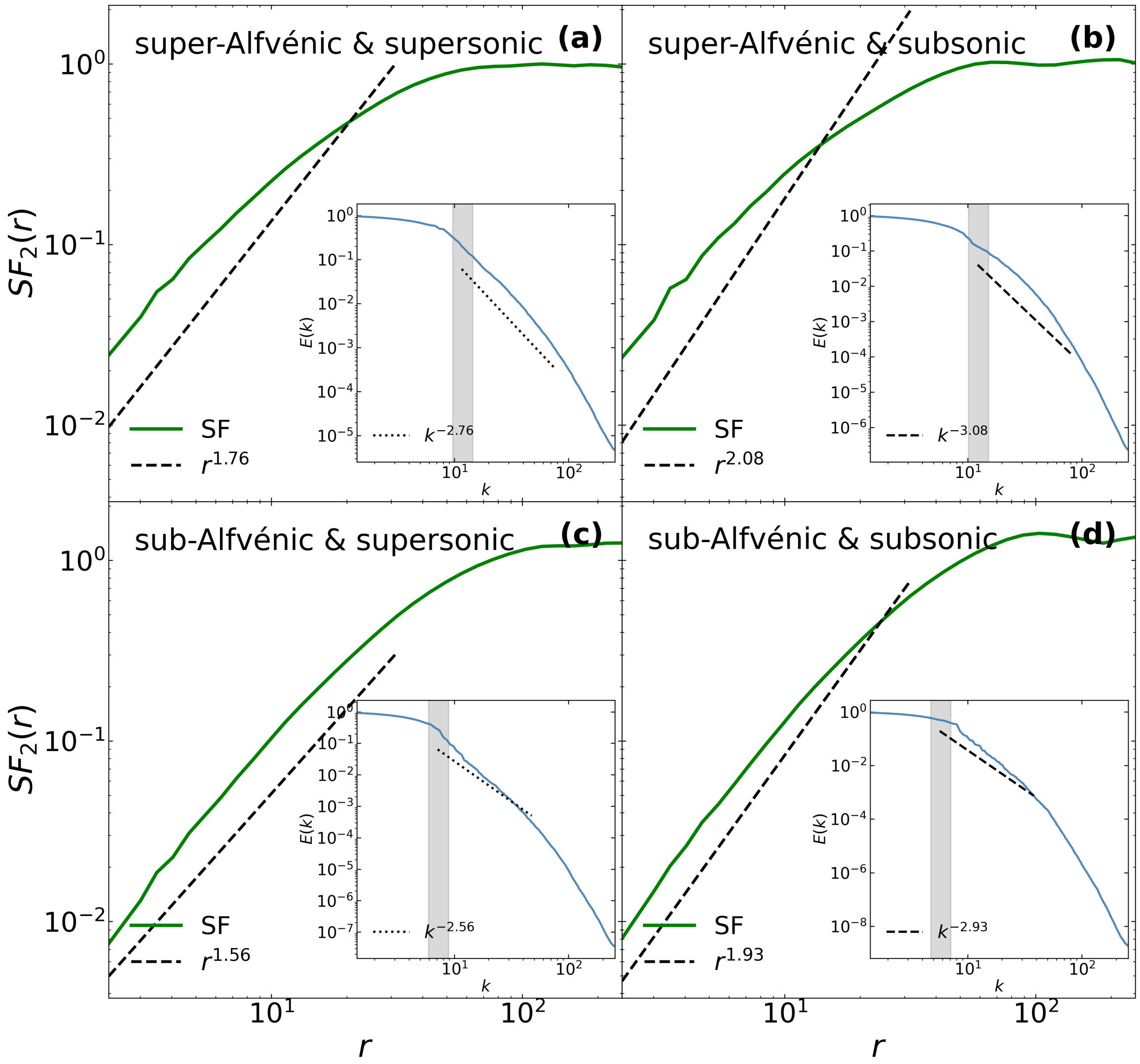

**Figure 9.** Polarization-intensity structure functions $\mathrm{SF}_2(r)$ (green solid lines) for the four turbulence regimes at $\nu = 10^4$ MHz. The black dashed lines indicate the reference slopes, i.e., $r^{5/3}$ for the supersonic cases and $r^2$ for the subsonic cases. Each panel contains an inset in the lower-right corner showing the corresponding power spectrum $E(k)$ (blue solid line), with the shaded band marking the weak-to-strong turbulence transition region and the black dotted lines indicating the reference slopes, i.e., $k^{-8/3}$ for the supersonic cases and $k^{-3}$ for the subsonic cases. All four panels share the same axis ranges.

## C. IMPACT OF MAGNETIC FIELD ORIENTATION ON POLARIZATION DEGREE

Fig. 10 visualizes the mean polarization degree $\langle p \rangle$ as a function of the mean magnetic field components $\langle B_\parallel \rangle$ and $\langle B_\perp \rangle$. Because the total mean field strength is fixed, the data points trace a quarter-circle arc. For super-Alfvénic turbulence (top panels), the mean polarization degree remains relatively low and exhibits only weak variations with magnetic field orientation. In contrast, sub-Alfvénic turbulence (bottom panels) generally exhibits a higher mean polarization degree and a stronger dependence on field orientation. Specifically, the polarization degree decreases as $\langle B_\parallel \rangle$ increases and $\langle B_\perp \rangle$ decreases. These results demonstrate that the polarization degree is more sensitive to magnetic field orientation in sub-Alfvénic turbulence than in super-Alfvénic turbulence.

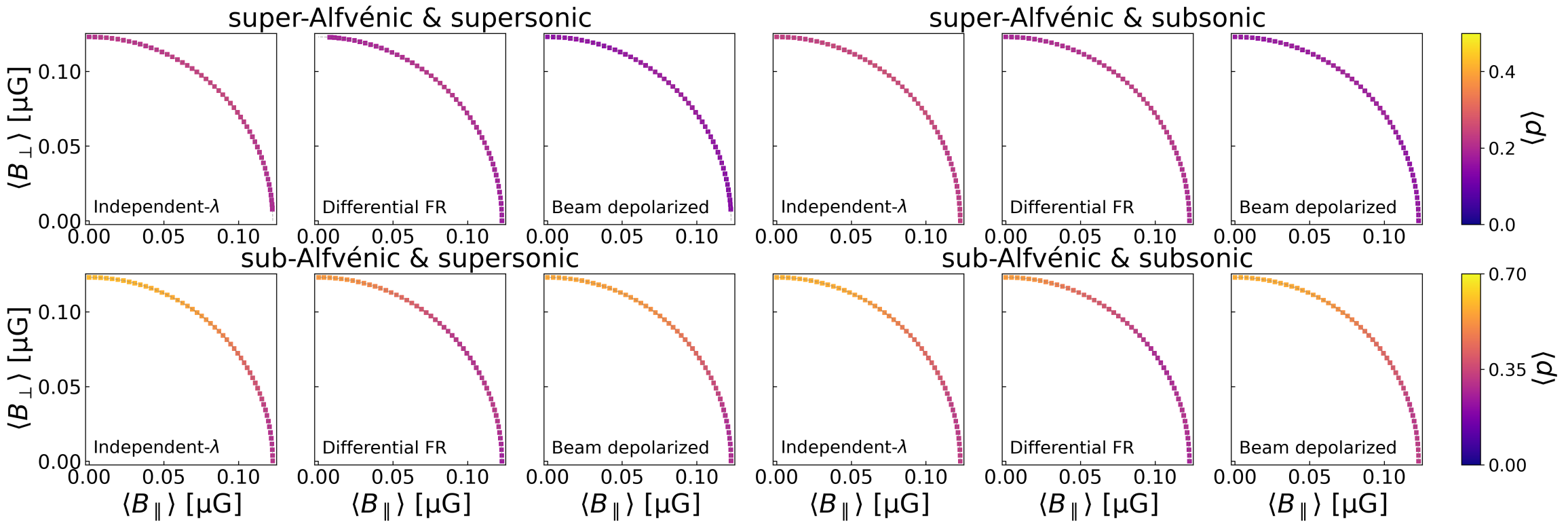

**Figure 10.** Heatmaps of the mean polarization degree $\langle p \rangle$ as a function of the mean parallel ($\langle B_\parallel \rangle$) and perpendicular ($\langle B_\perp \rangle$) magnetic field components. The top and bottom rows show super-Alfvénic and sub-Alfvénic turbulence, respectively, while the three columns correspond to independent-$\lambda$, differential Faraday rotation (Differential FR), and beam depolarization mechanisms. The color scale denotes the magnitude of $\langle p \rangle$, and the quarter-circle distributions reflect the fixed total mean magnetic field.